\documentclass[conference]{IEEEtran}
\PassOptionsToPackage{ruled, linesnumbered}{algorithm2e}
\usepackage[T1]{fontenc}
\usepackage[latin9]{inputenc}
\usepackage{color}
\usepackage{mathrsfs}
\usepackage{algorithm2e}
\usepackage{amsmath}
\usepackage{amssymb}
\usepackage{graphicx}

\makeatletter
\usepackage{cite, amsmath, amsfonts, amssymb, graphicx,lipsum}
\usepackage{amsthm}
\usepackage{balance}

\theoremstyle{remark}

\allowdisplaybreaks

\newcommand{\herm}{^{\mathsf{H}}}
\newcommand{\trans}{^{\mathsf{T}}}

\DeclareMathOperator{\tr}{tr}

\DeclareMathOperator{\diag}{\mathsf{diag}}
\DeclareMathOperator{\vecd}{\mathsf{vec}_{\mathsf{d}}}
\DeclareMathOperator{\maximize}{maximize}
\DeclareMathOperator{\st}{subject~to}

\usepackage{pgfplots}
\usepackage{pgfplotstable}
\usepackage{tikz}
\usetikzlibrary{pgfplots.groupplots}

\makeatother

\begin{document}
\title{\LARGE{Achievable Sum Secrecy Rate of STAR-RIS-Enabled MU-MIMO ISAC}}
\author{\IEEEauthorblockN{Ainara Kazymova\IEEEauthorrefmark{1}, Vaibhav Kumar\IEEEauthorrefmark{1}, Christina P\"opper\IEEEauthorrefmark{2}, and Marwa Chafii\IEEEauthorrefmark{1}\IEEEauthorrefmark{3}}\IEEEauthorblockA{\IEEEauthorrefmark{1}Engineering Division, New York University Abu Dhabi, UAE\\
\IEEEauthorrefmark{2}Science Division, New York University Abu Dhabi, UAE\\
\IEEEauthorrefmark{3}NYU WIRELESS, NYU Tandon School of Engineering, New York, USA \\
Email: ainara.kazymova@nyu.edu, vaibhav.kumar@ieee.org, \{christina.poepper, marwa.chafii\}@nyu.edu }}
\maketitle

{\let\thefootnote\relax\footnotetext{This work was supported by the Center for Cyber Security under the New York University Abu Dhabi Research Institute under Award G1104.}}

\begin{abstract}
Various emerging applications in sixth-generation (6G) wireless demand a seamless integration of communication and sensing services, driving the development of integrated sensing and communication (ISAC) systems. Using a common waveform for both functions introduces additional security challenges, as information-bearing signals are vulnerable to eavesdropping. While a variety of secure beamforming designs are proposed in literature for metasurface-enabled ISAC systems with single-antenna users and eavesdroppers, optimal designs for multi-antenna scenarios remain unexplored. This paper addresses this gap by studying a simultaneously transmitting and reflecting reconfigurable intelligent surface (STAR-RIS)-enabled multi-user multiple-input multiple-output (MU-MIMO) ISAC system and proposing an optimal beamforming design to maximize the sum secrecy rate while ensuring sensing quality for multiple targets. An alternating-optimization based iterative solution is developed to tackle the non-convex optimization problem.  The presented analysis confirms that the computational complexity of the proposed algorithm grows linearly with the number of metasurface elements. Our numerical results show the benefit of using STAR-RIS to increase the sum secrecy rate of the MU-MIMO ISAC system compared to its corresponding conventional RIS (cRIS)-enabled and w/o RIS systems.
\end{abstract}


\section{Introduction\protect\label{sec:Introduction}}

Given the increasing demand for seamless coexistence of communication and sensing services in various sixth-generation (6G) wireless applications, integrated sensing and communication (ISAC) has emerged as a cornerstone of research within the wireless community~\cite{24-Proc-IEEE-ISAC}. The advantages of intelligent metasurfaces in ISAC systems have also now been well-established~\cite{24-COMST-ISAC,25-MNET-zeRIS-ISAC}. However, since shared signals are utilized for both communication and sensing in the ISAC paradigm, these signals become potentially detectable by sensing targets. This introduces significant security risks, as sensing targets may potentially intercept confidential communication data. To address this security challenge, researchers are exploring numerous innovative physical-layer security approaches.

In this direction, the problem of ergodic secrecy rate maximization in a reconfigurable intelligent surface (RIS)-aided ISAC system, consisting of a single-antenna legitimate user, a single-antenna eavesdropper, and a target was considered in~\cite{24-TVT-RIS-ISAC-Xing}. The problem of radar signal-to-noise ratio (SNR) maximization in a RIS-enabled ISAC system with multiple single-antenna communication users, and one single-antenna eavesdropping target was considered in~\cite{24-TVT-MIMO-radar-SNR-maximization}. Similarly, the authors in~\cite{23-COMML-Secure-RIS-ISAC} considered a RIS-enabled ISAC system with multiple single-antenna communication users, one single-antenna eavesdropper, and a passive target, where they propose optimal beamforming design to minimize the maximum eavesdropping signal-to-interference-plus-noise ratio (SINR). The optimal beamforming design to maximize the sum secrecy rate in a RIS-enabled ISAC system with multiple single-antenna communication users, and one single-antenna eavesdropping target was proposed in~\cite{24-WCL-Secure-RIS-ISAC}. Similarly, optimal secure beamforming design for simultaneously transmitting and reflecting RIS (STAR-RIS)-enabled ISAC system with multiple single-antenna communication users and a single/multiple single-antenna eavesdropping target(s) were proposed in~\cite{24-TWC-STAR-RIS-NOMA-ISAC,24-IoTJ-STAR-RIS-IoE}. More recently, a joint secure and covert beamforming design for a STAR-RIS-enabled ISAC system was proposed with multiple single-antenna secrecy users, multiple single-antenna covert users (CUs), one single-antenna eavesdropping target, and one single-antenna warden to detect the existence of the wireless transmission from base station to CUs~\cite{25-IoTJ-SecureCovert}.

In summary, prior works proposed secure beamforming for metasurface-enabled ISAC systems with single-antenna users and eavesdroppers. However, the secrecy performance of \emph{MU-MIMO ISAC systems with multi-antenna eavesdroppers} remains unexplored. Given that future 6G devices (e.g., smartphones, wearables, IoT devices, autonomous vehicles) will feature multiple antennas, we study a STAR-RIS-aided MU-MIMO ISAC system where the base station and communication users have multiple antennas, and multiple single-antenna targets collaboratively act as eavesdroppers.

Against this background, the main contributions of this paper are:
\begin{enumerate}
\item[1)] We formulate the joint design of transmit, receive, and STAR-RIS beamformers to maximize the sum secrecy rate in a STAR-RIS-enabled MU-MIMO system, subject to sensing QoS constraints for multiple targets, a transmit power budget, STAR-RIS operation protocol, and unit-norm receive beamforming.
\item[2)] To ensure scalability, we propose an alternating optimization (AO)-based iterative algorithm for the non-convex problem, where receive beamformers are derived in closed-form, and transmit and STAR-RIS beamformers are obtained via a penalty dual decomposition-based accelerated gradient projection method. Complexity analysis of the proposed algorithm shows linear scaling with the number of metasurface elements.
\item[3)] Extensive simulations demonstrate the performance gains of the proposed STAR-RIS-enabled MU-MIMO ISAC system and highlight the impact of key design parameters. Results show superiority over benchmarks in terms of average sum secrecy rate, and reveal the effects of varying the number of user and eavesdropper antennas.
\end{enumerate}

\section{System Model and Problem Formulation}

\begin{figure}[b]
\begin{centering}
\includegraphics[width=0.65\columnwidth]{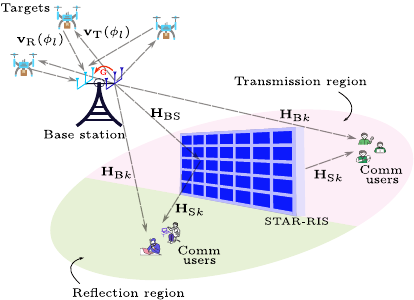}
\par\end{centering}
\caption{System model for STAR-RIS-enabled MU-MIMO ISAC.}

\label{fig:system_model}
\end{figure}
\paragraph*{System Model} 
We consider a STAR-RIS-enabled MU-MIMO ISAC system, as shown in Fig.~\ref{fig:system_model}, which consists of a dual-function radio-communication (DFRC) base station (BS), one STAR-RIS, $K$ communication users, and $L$ sensing targets. The communication users are denoted by $\mathrm{U}_{k}\ \forall k\in\mathcal{K}\triangleq\{1,2,\ldots,K\}$, and the targets are denoted by $\mathrm{E}_{l}\ \forall l\in\mathcal{L}\triangleq\{1,2,\ldots,L\}$. The BS has $N_{\mathrm{T}}$ transmit antennas and $N_{\mathrm{R}}$ receive antennas, STAR-RIS has $N_{\mathrm{S}}$ metasurface elements, $\mathrm{U}_{k}$ ($\forall k\in\mathcal{K}$) is equipped with $N_{\mathrm{U}}$ receive antennas, and $\mathrm{E}_{l}$ ($\forall l\in\mathcal{L}$) has a single receive antenna. We denote the wireless links between the transmit antennas array at the BS and STAR-RIS, transmit antennas array at the BS and $\mathrm{U}_{k}$, and STAR-RIS and $\mathrm{U}_{k}$ by $\mathbf{H}_{\mathrm{BS}}\in\mathbb{C}^{N_{\mathrm{S}}\times N_{\mathrm{T}}}$, $\mathbf{H}_{\mathrm{B}k}\in\mathbb{C}^{N_{\mathrm{U}}\times N_{\mathrm{T}}}$, and $\mathbf{H}_{\mathrm{S}k}\in\mathbb{C}^{N_{\mathrm{U}}\times N_{\mathrm{S}}}$, respectively. The steering vector from the transmit antenna array at the BS to $\mathrm{E}_{l}$ is denoted by $\mathbf{v}_{\mathrm{T}}(\phi_{l})\in\mathbb{C}^{1\times N_{\mathrm{T}}}$, and that from $\mathrm{E}_{l}$ to the receive antenna array of the BS is denoted by $\mathbf{v}_{\mathrm{R}}(\phi_{l})\in\mathbb{C}^{N_{\mathrm{R}}\times1}$, where $\phi_{l}$ is the azimuth angle of $\mathrm{E}_{l}$ w.r.t. the BS. The self-interference link between the transmit and receive antenna arrays at the BS is represented by $\mathbf{G}\in\mathbb{C}^{N_{\mathrm{R}}\times N_{\mathrm{T}}}$. Similar to~\cite{24-TVT-MIMO-radar-SNR-maximization,23-COMML-Secure-RIS-ISAC,24-WCL-Secure-RIS-ISAC} and~\cite{24-IoTJ-STAR-RIS-IoE}, we assume that the targets are located far away from the STAR-RIS, and therefore, the links between STAR-RIS and targets do not exist. The set of indexes of the communication users in the reflection and transmission regions of the STAR-RIS are respectively denoted by $\mathcal{K}_{\mathrm{R}}$ and $\mathcal{K}_{\mathrm{T}}$, where $\mathcal{K}_{\mathrm{R}}\cup\mathcal{K}_{\mathrm{T}}=\mathcal{K}\triangleq\{1,2,\ldots,K\}$ and $\mathcal{K}_{\mathrm{R}}\cap\mathcal{K}_{\mathrm{T}}=\emptyset$. In line with~\cite{24-TVT-MIMO-radar-SNR-maximization,23-COMML-Secure-RIS-ISAC,24-WCL-Secure-RIS-ISAC,24-TWC-STAR-RIS-NOMA-ISAC}, we assume that the BS has access to perfect instantaneous channel state information for all the links. However, different from the existing papers~\cite{24-TVT-RIS-ISAC-Xing,24-TVT-MIMO-radar-SNR-maximization,23-COMML-Secure-RIS-ISAC,24-WCL-Secure-RIS-ISAC,24-TWC-STAR-RIS-NOMA-ISAC,24-IoTJ-STAR-RIS-IoE}, where a single-antenna eavesdropper was considered, we assume that all the targets cooperatively eavesdrop to decode the information symbols intended for the communication users. In this manner, \emph{the targets effectively behave as a single eavesdropper with multiple antennas}, maximizing the chance of successfully decoding information symbols. Similar to~\cite{24-TWC-STAR-RIS-NOMA-ISAC}, it is assumed that the STAR-RIS operates in the power-splitting mode, i.e., for an incident signal $x$, the response of the $m^{\mathrm{th}}$ metasurface element of the STAR-RIS in the reflection and transmission regions are respectively modeled by $\theta_{m\mathrm{R}}x$ and $\theta_{m\mathrm{T}}x$, where $\sum_{\varkappa\in\{\mathrm{R},\mathrm{T}\}}|\theta_{m\varkappa}|^{2}=1$, and $\angle\theta_{m\varkappa}\in[0,2\pi)\ \forall m\in\mathscr{N}_{\mathrm{S}}\triangleq\{1,2,\ldots,N_{\mathrm{S}}\}$.

Following~\cite{24-TVT-MIMO-radar-SNR-maximization}, the transmitted signal from the BS is given by $\mathbf{x}=\sum\nolimits_{k\in\mathcal{K}}\mathbf{F}_{\mathrm{c}k}\mathbf{w}_{\mathrm{c}k}+\sum\nolimits_{l\in\mathcal{L}}\mathbf{F}_{\mathrm{s}l}\mathbf{w}_{\mathrm{s}l}$ where $\mathbf{w}_{\mathrm{c}k}\in\mathbb{C}^{N_{\min}\times1}$ is the signal vector intended for $\mathrm{U}_{k}$ with $\mathbf{F}_{\mathrm{c}k}\in\mathbb{C}^{N_{\mathrm{T}}\times N_{\min}}$ being the corresponding transmit precoding matrix, $N_{\min}\triangleq\min\{N_{\mathrm{T}},N_{\mathrm{U}}\}$, and $\mathbf{w}_{\mathrm{s}l}\in\mathbb{C}^{N_{\mathrm{T}}\times1}$ is the sensing vector corresponding to $\mathrm{E}_{l}$ with $\mathbf{F}_{\mathrm{s}l}\in\mathbb{C}^{N_{\mathrm{T}}\times N_{\mathrm{T}}}$ being the corresponding transmit precoding matrix. The signal received at $\mathrm{U}_{k}$ is then given by $\mathbf{y}_{\mathrm{U}_{k}}=(\mathbf{H}_{\mathrm{B}k}+\mathbf{H}_{\mathrm{S}k}\boldsymbol{\Theta}_{k}\mathbf{H}_{\mathrm{BS}})\mathbf{x}+\mathbf{n}_{\mathrm{U}_{k}}\triangleq\mathbf{Z}_{k}\mathbf{x}+\mathbf{n}_{\mathrm{U}_{k}}$, where $\mathbf{Z}_{k}=\mathbf{H}_{\mathrm{B}k}+\mathbf{H}_{\mathrm{S}k}\boldsymbol{\Theta}_{k}\mathbf{H}_{\mathrm{BS}}\in\mathbb{C}^{N_{\mathrm{U}}\times N_{\mathrm{T}}}$, and $\mathbf{n}_{\mathrm{U}_{k}}\sim\mathcal{CN}(\boldsymbol{0},\sigma_{\mathrm{U}_{k}}^{2}\mathbf{I})$ is the additive white Gaussian noise (AWGN) at $\mathrm{U}_{k}$. Moreover, $\boldsymbol{\Theta}_{k}=\diag(\boldsymbol{\theta}_{k})$, and 
\begin{equation}
\boldsymbol{\theta}_{k}=\begin{cases}
\boldsymbol{\theta}_{\mathrm{R}}=[\theta_{1\mathrm{R}},\ldots,\theta_{N_{\mathrm{S}}\mathrm{R}}]\trans\in\mathbb{C}^{N_{\mathrm{S}}\times1}, & \text{if}\ k\in\mathcal{K}_{\mathrm{R}}\\
\boldsymbol{\theta}_{\mathrm{T}}=[\theta_{1\mathrm{T}},\ldots,\theta_{N_{\mathrm{S}}\mathrm{T}}]\trans\in\mathbb{C}^{N_{\mathrm{S}}\times1}, & \text{otherwise}
\end{cases}.\label{eq:theta_definition}
\end{equation}

On the other hand, the signal received at $\mathrm{E}_{l}$ is given by $y_{\mathrm{E}_{l}}=\mathbf{v}_{\mathrm{T}}(\phi_{l})\mathbf{x}+n_{\mathrm{E}_{l}}$, where $n_{\mathrm{E}_{l}}\sim\mathcal{CN}(0,\sigma_{\mathrm{E}_{l}}^{2})$ is the AWGN at the target $\mathrm{E}_{l}$. Hence, the received signals at the targets can be collectively represented as $\mathbf{y}_{\mathrm{E}}=\mathbf{V}_{\mathrm{T}}\mathbf{x}+\mathbf{n}_{\mathrm{E}}$, where $\mathbf{V}_{\mathrm{T}}\triangleq[\mathbf{v}_{\mathrm{T}}(\phi_{1})\trans,\mathbf{v}_{\mathrm{T}}(\phi_{2})\trans,\ldots,\mathbf{v}_{\mathrm{T}}(\phi_{L})\trans]\trans\in\mathbb{C}^{L\times N_{\mathrm{T}}}$, and $\mathbf{n}_{\mathrm{E}}=[n_{\mathrm{E}_{1}},n_{\mathrm{E}_{2}},\ldots,n_{\mathrm{E}_{L}}]\trans$. Without loss of generality, we assume $\sigma_{\mathrm{U}_{k}}^{2}\ (\forall k\in\mathcal{K})=\sigma_{\mathrm{E}_{l}}^{2}\ (\forall l\in\mathcal{L})=\sigma^{2}$, and with a slight abuse of notations, we define $\mathbf{H}_{\mathrm{BS}}\leftarrow\mathbf{H}_{\mathrm{BS}}/\sigma$, $\mathbf{H}_{\mathrm{B}k}\leftarrow\mathbf{H}_{\mathrm{B}k}/\sigma$, and $\mathbf{v}_{\mathrm{T}}(\phi_{l})\leftarrow\mathbf{v}_{\mathrm{T}}(\phi_{l})/\sigma$.

Hence, the instantaneous achievable secrecy rate (in nats/s/Hz) for $\mathrm{U}_{k}$ is given by $R_{\mathrm{c}k}(\mathbf{J},\boldsymbol{\theta})=\ln\det\big[\mathbf{I}+\mathbf{Z}_{k}\mathbf{J}_{\mathrm{c}k}\mathbf{Z}_{k}\herm\mathbf{A}_{kk}^{-1}\big]-\ln\det\big[\mathbf{I}+\mathbf{V}_{\mathrm{T}}\mathbf{J}_{\mathrm{c}k}\mathbf{V}_{\mathrm{T}}\herm\mathbf{B}_{k}^{-1}\big]$, where $\mathbf{J}_{\mathrm{c}k}\triangleq\mathbf{F}_{\mathrm{c}k}\mathbf{F}_{\mathrm{c}k}\herm\ (\forall k\in\mathcal{K})$, $\mathbf{J}_{\mathrm{s}l}\triangleq\mathbf{F}_{\mathrm{s}l}\mathbf{F}_{\mathrm{s}l}\herm\ (\forall l\in\mathcal{L})$ are the \emph{transmit covariance matrices}, $\mathbf{J}\triangleq\{\mathbf{J}_{\mathrm{c}1},\mathbf{J}_{\mathrm{c}2},\ldots,\mathbf{J}_{\mathrm{c}K},\mathbf{J}_{\mathrm{\mathrm{s}1}},\mathbf{J}_{\mathrm{\mathrm{s}2}},\ldots,\mathbf{J}_{\mathrm{\mathrm{s}}L}\}\triangleq\{\bar{\mathbf{J}}_{1},\ldots,\bar{\mathbf{J}}_{K+L}\}$, $\boldsymbol{\Sigma}_{\mathrm{c}k}\triangleq\sum\nolimits_{\jmath\in\mathcal{K}\setminus\{k\}}\mathbf{J}_{\mathrm{c}\jmath}+\sum\nolimits_{l\in\mathcal{L}}\mathbf{J}_{\mathrm{s}l}$, $\mathbf{A}_{kk}\triangleq\mathbf{I}+\mathbf{Z}_{k}\boldsymbol{\Sigma}_{\mathrm{c}k}\mathbf{Z}_{k}\herm$, $\mathbf{B}_{k}\triangleq\mathbf{I}+\mathbf{V}_{\mathrm{T}}\boldsymbol{\Sigma}_{\mathrm{c}k}\mathbf{V}_{\mathrm{T}}\herm$, and $\boldsymbol{\theta}=[\boldsymbol{\theta}_{\mathrm{R}}\trans,\boldsymbol{\theta}_{\mathrm{T}}\trans]\trans$. At the same time, the echo signal received at the BS receive antenna array from the sensing targets can be given by $\mathbf{y}_{\mathrm{R}}=\sum\nolimits_{l\in\mathcal{L}}\alpha_{l}\mathbf{v}_{\mathrm{R}}(\phi_{l})\mathbf{v}_{\mathrm{T}}(\phi_{l})\mathbf{x}+\mathbf{G}\mathbf{x}+\mathbf{n}_{\mathrm{R}}\triangleq\sum_{l\in\mathcal{L}}\alpha_{l}\mathbf{V}_{\mathrm{R}l}\mathbf{x}+\mathbf{G}\mathbf{x}+\mathbf{n}_{\mathrm{R}}$, where $\mathbf{V}_{\mathrm{R}l}=\mathbf{v}_{\mathrm{R}}(\phi_{l})\mathbf{v}_{\mathrm{T}}(\phi_{l})$, $\mathbf{G}\leftarrow\mathbf{G}/\sigma$, $\mathbf{n}_{\mathrm{R}}\sim\mathcal{CN}(\boldsymbol{0},\mathbf{I})$ is the \emph{normalized} AWGN at the BS, and $\alpha_{l}$ depends on the target radar cross-section (RCS) of $\mathrm{E}_{l}$ with known $\mathbb{E}\{|\alpha_{l}|^{2}\}=\bar{\alpha}_{l}$. WLOG, we assume $\bar{\alpha}_{l}=\bar{\alpha}\ \forall l\in\mathcal{L}$. Before processing the received signal for sensing $\mathrm{E}_{l}$, the BS applies a receive beamforming vector $\boldsymbol{\varphi}_{l}\in\mathbb{C}^{N_{\mathrm{R}}\times1}$ resulting in the following post-combining received signal: $\boldsymbol{\varphi}_{l}\herm\mathbf{y}_{\mathrm{R}}=\sum\nolimits_{\ell\in\mathcal{L}}\alpha_{\ell}\boldsymbol{\varphi}_{l}\herm\mathbf{V}_{\mathrm{R}\ell}\mathbf{x}+\boldsymbol{\varphi}_{l}\herm\mathbf{G}\mathbf{x}+\boldsymbol{\varphi}_{l}\herm\mathbf{n}_{\mathrm{R}}.$ Hence, the post-combining SINR of the echo signal received at the BS from $\mathrm{E}_{l}$ is modeled as\footnote{We assume that all the targets are resolvable at the BS. The details of the underlying signal processing for resolving multiple targets at the BS is beyond the scope of this paper.} $\gamma_{\mathrm{s}l}(\mathbf{J},\boldsymbol{\varphi}_{l})=\bar{\alpha}\boldsymbol{\varphi}_{l}\herm\mathbf{V}_{\mathrm{R}l}\boldsymbol{\Sigma}\mathbf{V}_{\mathrm{R}l}\herm\boldsymbol{\varphi}_{l}\big[\bar{\alpha}\sum\nolimits_{\jmath\in\mathcal{L}\setminus\{l\}}\boldsymbol{\varphi}_{l}\herm\mathbf{V}_{\mathrm{R}\jmath}\boldsymbol{\Sigma}\mathbf{V}_{\mathrm{R}\jmath}\herm\boldsymbol{\varphi}_{l}+\boldsymbol{\varphi}_{l}\herm\mathbf{G}\boldsymbol{\Sigma}\mathbf{G}\herm\boldsymbol{\varphi}_{l}+\|\boldsymbol{\varphi}_{l}\herm\|^{2}\big]^{-1}$, where $\boldsymbol{\Sigma}\triangleq\sum\nolimits_{k\in\mathcal{K}}\mathbf{J}_{\mathrm{c}k}+\sum\nolimits_{l\in\mathcal{L}}\mathbf{J}_{\mathrm{s}l}$. One can therefore define the sensing rate of $\mathrm{E}_{l}$ at the receive antenna array of the BS as $R_{\mathrm{s}l}(\mathbf{J},\boldsymbol{\varphi}_{l})\triangleq\ln\big[1+\gamma_{\mathrm{s}l}(\mathbf{J},\boldsymbol{\varphi}_{l})\big]$. 

\paragraph*{Problem Formulation} 
With the given background, the problem of sum secrecy rate maximization for the communication users can be formulated as follows: 
{\small
\begin{subequations}
\label{eq:P1}
\begin{eqnarray}
(\mathbb{P}1) & \underset{\mathbf{J},\boldsymbol{\theta},\boldsymbol{\Phi}}{\maximize} & \sum\nolimits_{k\in\mathcal{K}}R_{\mathrm{c}k}(\mathbf{J},\boldsymbol{\theta}),\label{eq:obj-P1}\\
 & \st & R_{\mathrm{s}l}(\mathbf{J},\boldsymbol{\varphi}_{l})\geq\Delta_{\mathrm{s}l}\ \forall l\in\mathcal{L},\label{eq:sensing_rate-P1}\\
 &  & \tr(\boldsymbol{\Sigma})\leq P_{\max},\label{eq:TPC-P1}\\
 &  & |\theta_{m\mathrm{R}}|^{2}\!+\!|\theta_{m\mathrm{T}}|^{2}\!=\!1\ \forall m\in\mathscr{N}_{\mathrm{S}},\label{eq:UMC-P1}\\
 &  & \|\boldsymbol{\varphi}_{l}\|^{2}=1\ \forall l\in\mathcal{L}.\label{eq:UMC-rxFiler-P1}
\end{eqnarray}
\end{subequations}
}
In~$(\mathbb{P}1)$, the sensing QoS is guaranteed by~\eqref{eq:sensing_rate-P1} where $\Delta_{\mathrm{s}l}$ is the sensing rate threshold for $\mathrm{E}_{l}$,~\eqref{eq:TPC-P1} corresponds to the transmit power constraint with a budget on transmit power from the BS denoted by $P_{\max}$, and~\eqref{eq:UMC-P1} is due to the power-splitting protocol at the STAR-RIS. Moreover, $\boldsymbol{\Phi}\triangleq[\boldsymbol{\varphi}_{1},\boldsymbol{\varphi}_{2},\ldots,\boldsymbol{\varphi}_{L}]$, and~\eqref{eq:UMC-rxFiler-P1} ensures unit-norm receive beamforming at the BS receive antenna array for each receive beamforming vector $\boldsymbol{\varphi}_{l}$. One can easily note that~$(\mathbb{P}1)$ is non-convex due to~\eqref{eq:obj-P1},~\eqref{eq:sensing_rate-P1}, \eqref{eq:UMC-P1}, and~\eqref{eq:UMC-rxFiler-P1}. Furthermore, the difference-of-concave nature of the objective function, and the coupling between the optimization variables ($\mathbf{J}$ and $\boldsymbol{\theta}$) in~\eqref{eq:obj-P1}~\eqref{eq:sensing_rate-P1} makes~$(\mathbb{P}1)$ challenging to solve. 

\section{AO-Based Proposed Solution}

The size of~$(\mathbb{P}1)$ can be prohibitively large since STAR-RIS should be composed of hundreds of metasurface elements in practical deployments to unlock its real potential. Therefore, designing a low-complexity solution to~$(\mathbb{P}1)$ is crucial. Keeping this in mind, in this section, we propose a low-complexity algorithm to obtain a stationary solution to~$(\mathbb{P}1)$, and also present the corresponding complexity analysis. 

\subsection{Optimal $\boldsymbol{\Phi}$ for a given $(\mathbf{J},\boldsymbol{\theta})$}

For a given $(\mathbf{J},\boldsymbol{\theta})$, the design of $\boldsymbol{\Phi}$ depends only on~\eqref{eq:sensing_rate-P1} and~\eqref{eq:UMC-rxFiler-P1}. Using the notion of generalized Rayleigh quotient, eigenvalue decomposition (EVD), and~\eqref{eq:sensing_rate-P1}, the optimal $\boldsymbol{\varphi}_{l}\ (\forall l\in\mathcal{L})$ to maximize $\gamma_{\mathrm{s}l}(\mathbf{J},\boldsymbol{\varphi}_{l})$ is given by $\boldsymbol{\varphi}_{l,\mathrm{opt}}=\boldsymbol{\lambda}_{\max}\big(\boldsymbol{\Psi}_{2}^{-1}\boldsymbol{\Psi}_{1}\big)$, where $\boldsymbol{\Psi}_{1}\triangleq\bar{\alpha}\mathbf{V}_{\mathrm{R}l}\boldsymbol{\Sigma}\mathbf{V}_{\mathrm{R}l}\herm$, $\boldsymbol{\Psi}_{2}\triangleq\big(\mathbf{I}+\bar{\alpha}\sum\nolimits_{\jmath\in\mathcal{L}\setminus\{l\}}\mathbf{V}_{\mathrm{R}\jmath}\boldsymbol{\Sigma}\mathbf{V}_{\mathrm{R}\jmath}\herm+\mathbf{G}\boldsymbol{\Sigma}\mathbf{G}\herm\big)^{-1}$, and $\boldsymbol{\lambda}_{\max}(\mathbf{X})$ denotes the eigenvector corresponding to the largest eigenvalue of $\mathbf{X}$. It is noteworthy that the \emph{orthonormal} property of the eigenvectors (obtained via EVD) satisfies~\eqref{eq:UMC-rxFiler-P1} by default. 

\subsection{Optimal $(\mathbf{J},\boldsymbol{\theta})$ for a given $\boldsymbol{\Phi}$}

For a given $\boldsymbol{\Phi}$,~$(\mathbb{P}1)$ boils down to the following non-convex optimization problem:
\begin{equation}
(\mathbb{P}2):\qquad\underset{\mathbf{J},\boldsymbol{\theta}}{\maximize}\big\{\eqref{eq:obj-P1}|\eqref{eq:sensing_rate-P1},\eqref{eq:TPC-P1},\eqref{eq:UMC-P1}\big\}.\label{eq:P2}
\end{equation}
Note that the main challenge in solving~$(\mathbb{P}2)$ is due to the variable coupling in~\eqref{eq:obj-P1}, the sensing constraint in~\eqref{eq:sensing_rate-P1}, and the equality constraint~\eqref{eq:UMC-P1}. To tackle this issue, we first define $\mathscr{G}(\mathbf{J},\boldsymbol{\varphi}_{l},\Delta_{\mathrm{s}l},\tau_{l})\triangleq\Delta_{\mathrm{s}l}+\tau_{l}-R_{\mathrm{s}l}(\mathbf{J},\boldsymbol{\varphi}_{l})$. It is then easy to see that $\mathscr{G}(\mathbf{J},\boldsymbol{\varphi}_{l},\Delta_{\mathrm{s}l},\tau_{l})=0$ is equivalent to~\eqref{eq:sensing_rate-P1} for some $\tau_{l}\in\mathbb{R}_{+}$. By applying the penalty dual decomposition method~\cite{2020-PDD}, an \emph{augmented Lagrangian objective} function corresponding to~$(\mathbb{P}2)$ can be written as 
\begin{align}
 & \mathcal{R}_{\boldsymbol{\nu},\rho}(\mathbf{J},\boldsymbol{\theta},\boldsymbol{\tau})=\sum\nolimits_{k\in\mathcal{K}}R_{\mathrm{c}k}(\mathbf{J},\boldsymbol{\theta})\nonumber \\
 & -\!\!\sum_{l\in\mathcal{L}}\big[\nu_{l}\mathscr{G}(\mathbf{J},\boldsymbol{\varphi}_{l},\Delta_{\mathrm{s}l},\tau_{l})\!+\!0.5\rho^{-1}\mathscr{G}^{2}(\mathbf{J},\boldsymbol{\varphi}_{l},\Delta_{\mathrm{s}l},\tau_{l})\big],\label{eq:augmented_objective}
\end{align}
where $\boldsymbol{\tau}\triangleq[\tau_{1},\tau_{2},\ldots,\tau_{L}]\trans\in\mathbb{R}_{+}^{L\times1}$, $\boldsymbol{\nu}\triangleq[\nu_{1},\nu_{2},\ldots,\nu_{L}]\trans$ is the vector of Lagrange multipliers, and $\rho$ is the penalty multiplier. Therefore, using~\eqref{eq:augmented_objective}, the problem in~$(\mathbb{P}2)$ can be transformed to the following optimization problem: 
\begin{equation}
\!\!\!\!(\mathbb{P}3):\underset{\mathbf{J},\boldsymbol{\theta},\boldsymbol{\tau}}{\maximize}\big\{\mathcal{R}_{\boldsymbol{\nu},\rho}(\mathbf{J},\boldsymbol{\theta},\boldsymbol{\tau})|\boldsymbol{\tau}\in\mathbb{R}_{+}^{L\times1},\eqref{eq:TPC-P1},\eqref{eq:UMC-P1}\big\}.\!\!\!\!\label{eq:P3}
\end{equation}
One can note that the design variables are now decoupled in the constraints in~$(\mathbb{P}3)$. It can also be shown that the transformation of~$(\mathbb{P}2)$ to~$(\mathbb{P}3)$ does not change the optimality of the original problem. Since the objective function in~$(\mathbb{P}3)$ is smooth, and the constraints therein can be satisfied using projection operations, one can obtain numerically efficient solutions to the problem using Riemannian manifold optimization (RMO)-based approaches such as Conjugate-Gradient, Barzilai-Borwein, and Trust-Region methods. However, we use the accelerated proximal gradient method~\cite{15-APG}, due to its faster convergence, which is based on Euclidean gradients. It is noteworthy that although the proximal gradient method in~\cite{15-APG} was originally proposed for optimization problems with non-smooth objective function, it is also applicable for problems with smooth objective functions. 

Before proceeding further, we first obtain the gradients of the augmented objective function $\mathcal{R}_{\boldsymbol{\nu},\rho}(\mathbf{J},\boldsymbol{\theta},\boldsymbol{\tau})$ w.r.t. $\mathbf{J}$ and $\boldsymbol{\theta}$. It is straightforward to note that $\nabla_{\mathbf{J}}\mathcal{R}_{\boldsymbol{\nu},\rho}(\mathbf{J},\boldsymbol{\theta},\boldsymbol{\tau})=[\nabla_{\bar{\mathbf{J}}_{1}}\mathcal{R}_{\boldsymbol{\nu},\rho}(\mathbf{J},\boldsymbol{\theta},\boldsymbol{\tau}),\ldots,\nabla_{\bar{\mathbf{J}}_{K+L}}\mathcal{R}_{\boldsymbol{\nu},\rho}(\mathbf{J},\boldsymbol{\theta},\boldsymbol{\tau})]$. A closed-form expression for $\nabla_{\bar{\mathbf{J}}_{\varpi}}\mathcal{R}_{\boldsymbol{\nu},\rho}(\mathbf{J},\boldsymbol{\theta},\boldsymbol{\tau})\ \forall\varpi\in\{1,\ldots,K+L\}$ can be then given by $\nabla_{\bar{\mathbf{J}}_{\varpi}}\mathcal{R}_{\boldsymbol{\nu},\rho}(\mathbf{J},\boldsymbol{\theta},\boldsymbol{\tau})=\sum\nolimits_{k\in\mathcal{K}}\nabla_{\bar{\mathbf{J}}_{\varpi}}R_{\mathrm{c}k}(\mathbf{J},\boldsymbol{\theta})+\sum\nolimits_{l\in\mathcal{L}}\big[\nu_{l}+\rho^{-1}\mathscr{G}(\mathbf{J},\boldsymbol{\varphi}_{l},\Delta_{\mathrm{s}l},\tau_{l})\big]\nabla_{\bar{\mathbf{J}}_{\varpi}}R_{\mathrm{s}l}(\mathbf{J},\boldsymbol{\varphi}_{l})$, where 
\begin{equation}
\nabla_{\bar{\mathbf{J}}_{\varpi}}R_{\mathrm{c}k}(\mathbf{J},\boldsymbol{\theta})=\begin{cases}
\mathbf{Z}_{\varpi}\herm\mathbf{C}_{\varpi}^{-1}\mathbf{Z}_{\varpi}-\mathbf{V}_{\mathrm{T}}\herm\mathbf{D}^{-1}\mathbf{V}_{\mathrm{T}} & \text{if }\varpi=k\\
\mathbf{Z}_{k}\herm\big(\mathbf{C}_{k}^{-1}-\mathbf{A}_{kk}^{-1}\big)\mathbf{Z}_{k}\\
\qquad-\mathbf{V}_{\mathrm{T}}\herm\big(\mathbf{D}^{-1}-\mathbf{B}_{k}^{-1}\big)\mathbf{V}_{\mathrm{T}} & \text{otherwise},
\end{cases}\label{eq:grad_J_bar_comm_rate}
\end{equation}
 $\nabla_{\bar{\mathbf{J}}_{\varpi}}R_{\mathrm{s}l}(\mathbf{J},\boldsymbol{\varphi}_{l})$ is given by~\eqref{eq:grad_J_bar_sensing_rate}, shown on the next page, $\mathbf{C}_{\varpi}\triangleq\mathbf{I}+\mathbf{Z}_{k}\boldsymbol{\Sigma}\mathbf{Z}_{k}\herm$, and $\mathbf{D}\triangleq\mathbf{I}+\mathbf{V}_{\mathrm{T}}\boldsymbol{\Sigma}\mathbf{V}_{\mathrm{T}}\herm$. The closed-form expressions in~\eqref{eq:grad_J_bar_comm_rate} and~\eqref{eq:grad_J_bar_sensing_rate} are derived using the relation $\det[\mathbf{I}+\mathbf{X}\mathbf{Y}^{-1}]=\det[\mathbf{I}+\mathbf{Y}^{-1/2}\mathbf{X}\mathbf{Y}^{-1/2}]$, and~\cite{07-TSP_Differential}. Detailed derivation of the gradients is omitted due to space constraints. 
\begin{figure*}[t]
\begin{footnotesize}
\begin{multline}
\nabla_{\bar{\mathbf{J}}_{\varpi}}R_{\mathrm{s}l}(\mathbf{J},\boldsymbol{\varphi}_{l})=\frac{\bar{\alpha}\sum\nolimits_{\ell\in\mathcal{L}}\mathbf{V}_{\mathrm{R}\ell}\herm\boldsymbol{\varphi}_{l}\boldsymbol{\varphi}_{l}\herm\mathbf{V}_{\mathrm{R}\ell}+\mathbf{G}\herm\boldsymbol{\varphi}_{l}\boldsymbol{\varphi}_{l}\herm\mathbf{G}}{\bar{\alpha}\sum\nolimits_{\ell\in\mathcal{L}}\boldsymbol{\varphi}_{l}\herm\mathbf{V}_{\mathrm{R}\ell}\boldsymbol{\Sigma}\mathbf{V}_{\mathrm{R}\ell}\herm\boldsymbol{\varphi}_{l}+\boldsymbol{\varphi}_{l}\herm\mathbf{G}\boldsymbol{\Sigma}\mathbf{G}\herm\boldsymbol{\varphi}_{l}+\|\boldsymbol{\varphi}_{l}\herm\|^{2}}
-\frac{\bar{\alpha}\sum\nolimits_{\jmath\in\mathcal{L}\setminus\{l\}}\mathbf{V}_{\mathrm{R}\jmath}\herm\boldsymbol{\varphi}_{l}\boldsymbol{\varphi}_{l}\herm\mathbf{V}_{\mathrm{R}\jmath}+\mathbf{G}\herm\boldsymbol{\varphi}_{l}\boldsymbol{\varphi}_{l}\herm\mathbf{G}}{\bar{\alpha}\sum\nolimits_{\jmath\in\mathcal{L}\setminus\{l\}}\boldsymbol{\varphi}_{l}\herm\mathbf{V}_{\mathrm{R}\jmath}\boldsymbol{\Sigma}\mathbf{V}_{\mathrm{R}\jmath}\herm\boldsymbol{\varphi}_{l}+\boldsymbol{\varphi}_{l}\herm\mathbf{G}\boldsymbol{\Sigma}\mathbf{G}\herm\boldsymbol{\varphi}_{l}+\|\boldsymbol{\varphi}_{l}\herm\|^{2}}.\label{eq:grad_J_bar_sensing_rate}
\end{multline}
\end{footnotesize}
\rule{1\textwidth}{1pt}
\end{figure*}
\begin{figure*}[t]
\begin{minipage}[]{0.24\textwidth}%
\begin{center}
\includegraphics[width=0.99\columnwidth,totalheight=3.5cm]{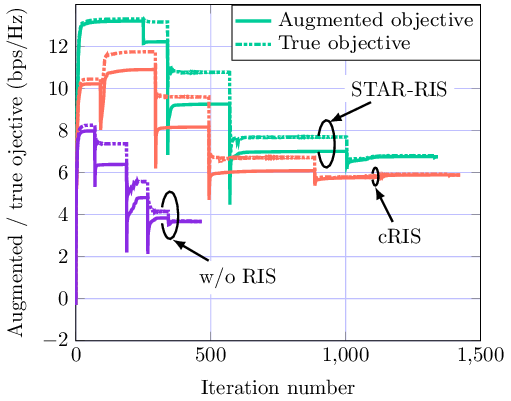}\vspace{-0.08in}
\par\end{center}
\caption{Comparison of convergence behavior for different MU-MIMO ISAC systems.}

\label{fig:convergence}%
\end{minipage}\hfill{}%
\begin{minipage}[]{0.24\textwidth}%
\begin{center}
\includegraphics[width=0.99\columnwidth,totalheight=3.5cm]{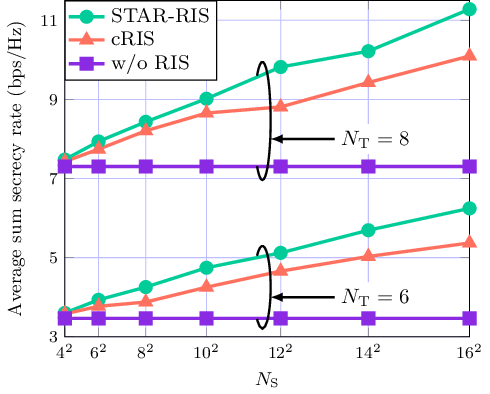}\vspace{-0.08in}
\par\end{center}
\caption{Impact of the number of metasurface elements ($N_{\mathrm{S}}$) on the secrecy rate.}
\label{fig:vary_Ns}%
\end{minipage}\hfill{}%
\begin{minipage}[]{0.24\textwidth}%
\begin{center}
\includegraphics[width=0.99\columnwidth,totalheight=3.5cm]{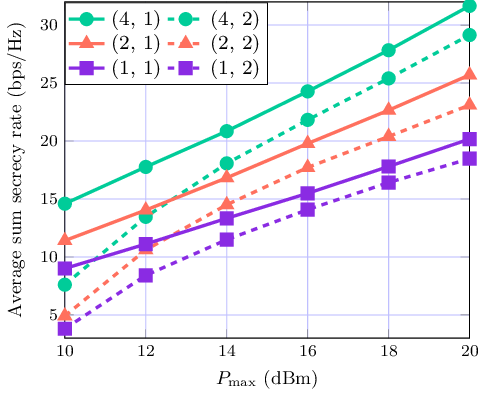}\vspace{-0.08in}
\par\end{center}
\caption{Impact of transmit power budget on the secrecy rate of STAR-RIS-enabled system.}
\label{fig:vary_Pmax}%
\end{minipage}\hfill{}%
\begin{minipage}[]{0.24\textwidth}%
\begin{center}
\includegraphics[width=0.99\columnwidth,totalheight=3.5cm]{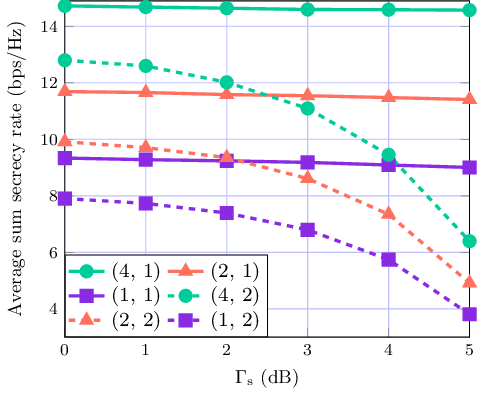}\vspace{-0.08in}
\par\end{center}
\caption{Impact of the sensing SINR threshold on the secrecy rate of STAR-RIS-enabled system.}
\label{fig:vary_GammaS}%
\end{minipage}

\end{figure*}
Similarly, an analytical expression for $\nabla_{\boldsymbol{\theta}_{\varkappa}}\mathcal{R}_{\boldsymbol{\nu},\rho}(\mathbf{J},\boldsymbol{\theta},\boldsymbol{\tau})\ \forall\varkappa\in\{\mathrm{R},\mathrm{T}\}$ is given by $\nabla_{\boldsymbol{\theta}_{\varkappa}}\mathcal{R}_{\boldsymbol{\nu},\rho}(\mathbf{J},\boldsymbol{\theta},\boldsymbol{\tau})=\sum\nolimits_{k\in\mathcal{K}}\nabla_{\boldsymbol{\theta}_{\varkappa}}R_{\mathrm{c}k}(\mathbf{J},\boldsymbol{\theta})$, where 
\begin{small}
\begin{equation}
\nabla_{\boldsymbol{\theta}_{\varkappa}}R_{kt}(\mathbf{J},\boldsymbol{\theta})=\begin{cases}
\vecd\big\{\mathbf{H}_{\mathrm{S}k}\herm\big(\mathbf{C}_{k}^{-1}\mathbf{Z}_{k}\boldsymbol{\Sigma}\\
\quad-\mathbf{A}_{kk}^{-1}\mathbf{Z}_{k}\boldsymbol{\Sigma}_{\mathrm{c}k}\big)\mathbf{H}_{\mathrm{BS}}\herm\big\} & \text{if }k\in\mathcal{K}_{\varkappa}\\
\boldsymbol{0} & \text{otherwise}.
\end{cases}\label{eq:grad_theta_comm_rate}
\end{equation}
\end{small}
Here we use the fact that $\mathscr{G}(\mathbf{J},\boldsymbol{\varphi}_{l},\Delta_{\mathrm{s}l},\tau_{l})$ is not a function of $\boldsymbol{\theta}_{\varkappa}$, and~\eqref{eq:grad_theta_comm_rate} is obtained using~\cite{07-TSP_Differential} and the properties of patterned matrices. We finally define $\nabla_{\boldsymbol{\theta}}\mathcal{R}_{\boldsymbol{\nu},\rho}(\mathbf{J},\boldsymbol{\theta},\boldsymbol{\tau})=\big[\big(\nabla_{\boldsymbol{\theta}_{\mathrm{R}}}\mathcal{R}_{\boldsymbol{\nu},\rho}(\mathbf{J},\boldsymbol{\theta},\boldsymbol{\tau})\big)\trans,\nabla_{\boldsymbol{\theta}_{\mathrm{T}}}\mathcal{R}_{\boldsymbol{\nu},\rho}(\mathbf{J},\boldsymbol{\theta},\boldsymbol{\tau})\big)\trans\big]\trans$. Being equipped with the closed-form expressions of $\nabla_{\mathbf{J}}\mathcal{R}_{\boldsymbol{\nu},\rho}(\mathbf{J},\boldsymbol{\theta},\boldsymbol{\tau})$ and $\nabla_{\boldsymbol{\theta}}\mathcal{R}_{\boldsymbol{\nu},\rho}(\mathbf{J},\boldsymbol{\theta},\boldsymbol{\tau})$, we now propose an AO-based iterative routine, given in~\textbf{Algorithm~\ref{algorithm-2}}, to obtain a stationary solution to~$(\mathbb{P}1)$. 

\begin{algorithm}[b!]
\begin{footnotesize}
\caption{AO-Based Algorithm to Solve~$(\mathbb{P}1)$ for fixed $\boldsymbol{\nu}$ and $\rho$}

\label{algorithm-1}

\KwIn{$\mathbf{J}^{(0)}$, $\boldsymbol{\theta}^{(0)}$, $\boldsymbol{\tau}^{(1)}$, $\boldsymbol{\nu}$, $\rho$, $t^{(0)}\!=\!t^{(1)}\!=\!1$, $\zeta\leq1$}

$\mathbf{J}^{(1)}=\mathbf{M}^{(1)}=\mathbf{J}^{(0)}$, $\boldsymbol{\theta}^{(1)}=\boldsymbol{\xi}^{(1)}=\boldsymbol{\theta}^{(0)}$\;

\For{$\imath=1,2,\ldots$}{

\tcc{\textcolor{brown}{Update transmit precoders $\mathbf{J}$}}

$\mathbf{Q}^{(\imath)}=\mathbf{J}^{(\jmath)}\!+\!\frac{t^{(\imath-1)}}{t^{(\imath)}}\big\{\mathbf{M}^{(\imath)}\!-\!\mathbf{J}^{(\imath)}\big\}+\frac{t^{(\imath-1)}-1}{t^{(\imath)}}\big\{\mathbf{J}^{(\imath)}-\mathbf{J}^{(\imath-1)}\big\}$\;

$\mathbf{M}^{(\imath+1)}=\Pi_{\mathcal{J}}\big\{\mathbf{Q}^{(\imath)}\!+\mu_{1}\nabla_{\mathbf{Q}}\mathcal{R}_{\boldsymbol{\nu},\rho}(\mathbf{Q}^{(\imath)},\boldsymbol{\theta}^{(\imath)},\boldsymbol{\tau}^{(\imath)})\big\}$\;

$\mathbf{U}^{(\imath+1)}=\Pi_{\mathcal{J}}\big\{\mathbf{J}^{(\imath)}+\mu_{2}\nabla_{\mathbf{J}}\mathcal{R}_{\boldsymbol{\nu},\rho}(\mathbf{J}^{(\imath)},\boldsymbol{\theta}^{(\imath)},\boldsymbol{\tau}^{(\imath)})\big\}$\;

$\mathbf{J}^{(\imath+1)}=\begin{cases}
\mathbf{M}^{(\imath+1)} & \text{if\ }\mathcal{R}_{\boldsymbol{\nu},\rho}(\mathbf{M}^{(\imath+1)},\boldsymbol{\theta}^{(\imath)},\boldsymbol{\tau}^{(\imath)})\\
 & \qquad\geq\mathcal{R}_{\boldsymbol{\nu},\rho}(\mathbf{U}^{(\imath+1)},\boldsymbol{\theta}^{(\imath)},\boldsymbol{\tau}^{(\imath)})\\
\mathbf{U}^{(\imath+1)} & \text{otherwise}
\end{cases}$

\tcc{\textcolor{brown}{Update RIS beamformers $\boldsymbol{\theta}$}}

$\boldsymbol{\omega}^{(\imath)}=\boldsymbol{\theta}^{(\imath)}\!+\!\frac{t^{(\imath-1)}}{t^{(\imath)}}\big\{\boldsymbol{\xi}^{(\imath)}\!-\!\boldsymbol{\theta}^{(\imath)}\big\}+\frac{t^{(\imath-1)}-1}{t^{(\imath)}}\big\{\boldsymbol{\theta}^{(\imath)}-\boldsymbol{\theta}^{(\imath-1)}\big\}$\;

$\boldsymbol{\xi}^{(\imath+1)}\!=\!\Pi_{\vartheta}\big\{\boldsymbol{\omega}^{(\imath)}+\eta_{1}\nabla_{\boldsymbol{\omega}}\mathcal{R}_{\boldsymbol{\nu},\rho}\big(\mathbf{J}^{(\imath+1)},\boldsymbol{\omega}^{(\imath)},\boldsymbol{\tau}^{(\imath)}\big)\big\}$\;

$\boldsymbol{\wp}^{(\imath+1)}=\Pi_{\vartheta}\big\{\boldsymbol{\theta}^{(\imath)}+\eta_{2}\nabla_{\boldsymbol{\theta}}\mathcal{R}_{\boldsymbol{\nu},\rho}\big(\mathbf{J}^{(\imath+1)},\boldsymbol{\theta}^{(\imath)},\boldsymbol{\tau}^{(\imath)}\big)\big\}$\;

$\boldsymbol{\theta}^{(\imath+1)}=\begin{cases}
\boldsymbol{\xi}^{(\imath+1)} & \text{if\ }\mathcal{R}_{\boldsymbol{\nu},\rho}(\mathbf{J}^{(\imath+1)},\boldsymbol{\xi}^{(\imath+1)},\boldsymbol{\tau}^{(\imath)})\\
 & \qquad\geq\mathcal{R}_{\boldsymbol{\nu},\rho}(\mathbf{J}^{(\imath+1)},\boldsymbol{\wp}^{(\imath+1)},\boldsymbol{\tau}^{(\imath)})\\
\boldsymbol{\wp}^{(\imath+1)} & \text{otherwise}
\end{cases}$

\tcc{\textcolor{brown}{Update $\boldsymbol{\tau}$}}

$\tau_{l}^{(\imath+1)}\!=\!\max\big\{0,R_{\mathrm{s}l}(\mathbf{J}^{(\imath+1)},\boldsymbol{\varphi}_{l}^{(\imath)})\!-\!\Delta_{\mathrm{s}l}\!-\!\nu_{l}\rho\big\}\forall l\in\mathcal{L}$\;

\tcc{\textcolor{brown}{Update receive beamformers $\boldsymbol{\Phi}$}}

$\boldsymbol{\varphi}_{l}^{(\imath+1)}=\boldsymbol{\lambda}_{\max}\big((\boldsymbol{\Psi}_{2}^{(\imath)})^{-1}\boldsymbol{\Psi}_{1}^{(\imath)}\big)\forall l\in\mathcal{L}$\;

\tcc{\textcolor{brown}{Update $t$}}

$t^{(\imath+1)}=0.5\big\{1+\sqrt{1+4(t^{(\imath)})^{2}}\big\}$\;

}

$\mathbf{J}^{(0)}\leftarrow\mathbf{J}^{(\imath+1)}$, $\boldsymbol{\theta}^{(0)}\leftarrow\boldsymbol{\theta}^{(\imath+1)}$, $\boldsymbol{\Phi}^{(0)}\leftarrow\boldsymbol{\Phi}^{(\imath+1)}$\;

\KwOut{$\mathbf{J}^{(0)}$, $\boldsymbol{\theta}^{(0)}$, $\boldsymbol{\Phi}^{(0)}$}
\end{footnotesize}
\end{algorithm}

\begin{algorithm}[t]
\begin{footnotesize}
\caption{AO-Based Algorithm to Solve~$(\mathbb{P}1)$}

\label{algorithm-2}

\KwIn{$\mathbf{J}^{(0)}$, $\boldsymbol{\theta}^{(0)}$, $\boldsymbol{\tau}^{(1)}$, $\boldsymbol{\nu}$, $\rho$, $\zeta\leq1$}

\Repeat{convergence}{

\tcc{\textcolor{brown}{Obtain optimal $(\mathbf{J}^{(0)},\boldsymbol{\theta}^{(0)},\boldsymbol{\Phi}^{(0)})$}}

Solve~$(\mathbb{P}1)$ for fixed $(\boldsymbol{\nu},\rho)$ using \textbf{Algorithm~\ref{algorithm-1}}

\tcc{\textcolor{brown}{Update Lagrange multipliers $\boldsymbol{\nu}$}}

$\nu_{l}\leftarrow\nu_{l}+\rho^{-1}\mathscr{G}(\mathbf{J}^{(0)},\boldsymbol{\varphi}_{l}^{(0)},\Delta_{\mathrm{s}l},\tau_{l}^{(0)})\forall l\in\mathcal{L}$\;

\tcc{\textcolor{brown}{Update penalty multiplier} $\rho$}

$\rho\leftarrow\zeta\rho$\;

}

\tcc{\textcolor{brown}{Final assignment}}

$\mathbf{J}_{\mathrm{opt}}\leftarrow\mathbf{J}^{(0)}$, $\boldsymbol{\theta}_{\mathrm{opt}}\leftarrow\boldsymbol{\theta}^{(0)}$, $\boldsymbol{\Phi}_{\mathrm{opt}}\leftarrow\boldsymbol{\Phi}^{(0)}$\;

\KwOut{$\mathbf{J}_{\mathrm{opt}}$, $\boldsymbol{\theta}_{\mathrm{opt}}$, $\boldsymbol{\Phi}_{\mathrm{opt}}$}
\end{footnotesize}
\end{algorithm}

In~\textbf{Algorithm~\ref{algorithm-2}}, we first obtain the optimal $(\mathbf{J},\boldsymbol{\theta},\boldsymbol{\tau})$ using~\textbf{Algorithm~\ref{algorithm-1}} for a given $\boldsymbol{\tau}$ and $\rho$. More specifically, in~\textbf{Algorithm~\ref{algorithm-1}}, we start with $\mathbf{J}^{(0)}=[\boldsymbol{0},\ldots,\boldsymbol{0}]$, and $\boldsymbol{\theta}_{\mathrm{R}}^{(0)}=\boldsymbol{\theta}_{\mathrm{T}}^{(0)}=\sqrt{0.5}[1,\ldots,1]\trans$. Note that these values satisfy~\eqref{eq:TPC-P1} and~\eqref{eq:UMC-P1}, respectively. Similarly, we choose a random initial value of $\boldsymbol{\varphi}_{l}\ (\forall l\in\mathcal{L})$ such that~\eqref{eq:UMC-rxFiler-P1} is satisfied. We then update $\mathbf{J}$ using the following procedure: In step~3 of~\textbf{Algorithm~\ref{algorithm-1}}, we find an extrapolated point $\mathbf{Q}$, and then in step 4, we move along the gradient direction $\nabla_{\mathbf{Q}}\mathcal{R}_{\boldsymbol{\nu},\rho}(\mathbf{Q},\boldsymbol{\theta},\boldsymbol{\tau})$ with a step-size $\mu_{1}$. A closed-form expression for $\nabla_{\mathbf{Q}}\mathcal{R}_{\boldsymbol{\nu},\rho}(\mathbf{Q},\boldsymbol{\theta},\boldsymbol{\tau})$ is obtained in a way similar to $\nabla_{\mathbf{J}}\mathcal{R}_{\boldsymbol{\nu},\rho}(\mathbf{J},\boldsymbol{\theta},\boldsymbol{\tau})$. Moreover, the feasible set of $\mathbf{J}$, denoted by $\mathcal{J}$ is defined as follows: $\mathcal{J}\triangleq\big\{\big[\bar{\mathbf{J}}_{1},\ldots\bar{\mathbf{J}}_{\varpi},\ldots,\bar{\mathbf{J}}_{K+L}\big],\in\mathbb{C}^{(K+L)\times N_{\mathrm{T}}\times N_{\mathrm{T}}}:\tr\big(\sum\nolimits_{\varpi}\bar{\mathbf{J}}_{\varpi}\big)\leq P_{\max};\bar{\mathbf{J}}_{\varpi}\succeq\boldsymbol{0}\big\}$. By letting $\widetilde{\mathbf{Q}}\triangleq\mathbf{Q}\!+\mu_{1}\nabla_{\mathbf{Q}}\mathcal{R}_{\boldsymbol{\nu},\rho}(\mathbf{Q},\boldsymbol{\theta},\boldsymbol{\tau})$, the projection of $\widetilde{\mathbf{Q}}$ onto $\mathcal{J}$, i.e., $\Pi_{\mathcal{J}}(\widetilde{\mathbf{Q}})$ can be obtained using the well-known \emph{water-filling algorithm}. A similar routine is followed in step~5 of~\textbf{Algorithm~\ref{algorithm-1}}. However, since $\mathbf{Q}$ can be a bad extrapolation of $\mathbf{J}$, we employ a monitor function in step~6 of the algorithm to finally obtain an optimal value of $\mathbf{J}$ for the current iteration. To find the optimal STAR-RIS beamforming $\boldsymbol{\theta}$ for the current iteration, we follow the series of steps in 7\textendash 10, which are similar to those in 3\textendash 6 in~\textbf{Algorithm~\ref{algorithm-1}}. The feasible set of $\boldsymbol{\theta}$, denoted by $\vartheta$, is defined as follows: $\vartheta\triangleq\big\{\big[\boldsymbol{\theta}_{\mathrm{R}}\trans,\boldsymbol{\theta}_{\mathrm{T}}\trans\big]\trans\in\mathbb{C}^{2N_{\mathrm{S}}\times1}:\eqref{eq:UMC-P1}\big\}$.  Define $\widetilde{\boldsymbol{\omega}}\triangleq[\widetilde{\omega}_{1\mathrm{R}},\ldots\widetilde{\omega}_{N_{\mathrm{S}}\mathrm{R}},\widetilde{\omega}_{1\mathrm{T}},\ldots\widetilde{\omega}_{N_{\mathrm{S}}\mathrm{T}}]\trans=\boldsymbol{\omega}+\eta_{1}\nabla_{\boldsymbol{\omega}}\mathcal{R}_{\boldsymbol{\nu},\rho}(\mathbf{J},\boldsymbol{\omega},\boldsymbol{\tau})$, then the projection of $\widetilde{\boldsymbol{\omega}}$ onto the feasible set $\vartheta$ is given by $\Pi_{\vartheta}(\widetilde{\boldsymbol{\omega}})=\boldsymbol{\xi}=[\xi_{1\mathrm{R}},\ldots,\xi_{N_{\mathrm{S}}\mathrm{R}},\xi_{1\mathrm{T}},\ldots,\xi_{N_{\mathrm{S}}\mathrm{T}}]\trans$, where $\xi_{m\mathrm{\mathrm{R}}}=\frac{\widetilde{\omega}_{m\mathrm{R}}}{\sqrt{|\widetilde{\omega}_{m\mathrm{R}}|^{2}+|\widetilde{\omega}_{m\mathrm{T}}|^{2}}}$ and $\xi_{m\mathrm{\mathrm{T}}}=\frac{\widetilde{\omega}_{m\mathrm{T}}}{\sqrt{|\widetilde{\omega}_{m\mathrm{R}}|^{2}+|\widetilde{\omega}_{m\mathrm{T}}|^{2}}},\ \forall m\in\mathscr{N}_{\mathrm{S}}.$ For the case when $|\widetilde{\omega}_{m\mathrm{R}}|^{2}+|\widetilde{\omega}_{m\mathrm{T}}|^{2}=0\ \exists m\in\mathscr{N}_{\mathrm{S}}$, we choose $\xi_{m\mathrm{\mathrm{R}}}$ and $\xi_{m\mathrm{\mathrm{T}}}$ randomly, such that $|\xi_{m\mathrm{\mathrm{R}}}|^{2}+|\xi_{m\mathrm{\mathrm{T}}}|^{2}=1$. Note that the optimal value of $\mu_{1}$, $\mu_{2}$, $\eta_{1}$, and $\eta_{2}$, in steps 4, 5, 8, and 9, respectively, is obtained using a \emph{line search procedure}, following the Armijo\textendash Goldstein condition~\cite{Armijo-Goldstein-Condition}. In the sequel, we update $\boldsymbol{\tau}$ in step~11, the receive beamformers $\boldsymbol{\Phi}$ in step~12, and $t$ in step~13. The steps 3\textendash 13 in~\textbf{Algorithm~\ref{algorithm-1}} are repeated until the relative change in the augmented objective function is below a predefined convergence threshold. 

After obtaining the optimal $(\mathbf{J},\boldsymbol{\theta},\boldsymbol{\Phi})$ in step~2 of~\textbf{Algorithm~\ref{algorithm-2}} for the given iteration, we update the Lagrange multipliers, $\boldsymbol{\nu}$, and the penalty multiplier, $\rho$, in steps 3 and 4, respectively, in~\textbf{Algorithm~\ref{algorithm-2}}. We repeat the steps 2\textendash 4 in~\textbf{Algorithm~\ref{algorithm-2}} until the relative change between the augmented objective, i.e., $\mathcal{R}_{\boldsymbol{\nu},\rho}(\mathbf{J},\boldsymbol{\theta},\boldsymbol{\tau})$, and the true objective, i.e.,$\sum_{k\in\mathcal{K}}R_{\mathrm{c}k}(\mathbf{J},\boldsymbol{\theta})$ becomes less than a predefined convergence threshold.

\subsection{Complexity Analysis}

We present the \emph{per iteration computational complexity} of~\textbf{Algorithm~\ref{algorithm-2}} in terms of the number of complex-valued matrix multiplications. It is easy to note that the complexity of~\textbf{Algorithm~\ref{algorithm-2}} is dominated by that of~\textbf{Algorithm~\ref{algorithm-1}}. Note that in~\textbf{Algorithm~\ref{algorithm-1}},\textbf{ }the complexity of computing $[\mathbf{Z}_{1},\mathbf{Z}_{2},\ldots,\mathbf{Z}_{K}]$ is given by $\mathcal{O}\{KN_{\mathrm{T}}N_{\mathrm{S}}N_{\mathrm{U}}\}$. Next, the complexity of computing $\nabla_{\mathbf{Q}}\mathcal{R}_{\boldsymbol{\nu},\rho}(\mathbf{Q},\boldsymbol{\theta},\boldsymbol{\tau})$ and $\nabla_{\mathbf{J}}\mathcal{R}_{\boldsymbol{\nu},\rho}(\mathbf{J},\boldsymbol{\theta},\boldsymbol{\tau})$ can be represented by $\mathcal{O}\{KN_{\mathrm{T}}N_{\mathrm{S}}N_{\mathrm{U}}+(K+L)[KN_{\mathrm{U}}^{3}+KN_{\mathrm{T}}^{2}N_{\mathrm{U}}+KN_{\mathrm{T}}N_{\mathrm{U}}^{2}+LN_{\mathrm{T}}^{2}N_{\mathrm{R}}]\}$. At the same time, the complexity of projection onto $\mathcal{J}$ using water-filling algorithm in steps 4 and 5 of~\textbf{Algorithm~\ref{algorithm-1}} leads to a computational complexity of $\mathcal{O}\{(K+L)N_{\mathrm{T}}^{3}\}$. Similarly, the computational complexity of steps 8 and 9 is given by $\mathcal{O}\{2KN_{\mathrm{U}}N_{\mathrm{T}}^{2}+2KN_{\mathrm{U}}^{2}N_{\mathrm{T}}+KN_{\mathrm{U}}N_{\mathrm{T}}N_{\mathrm{S}}\}$. It is noteworthy that the complexity associated with steps 3, 6, 7, and 10\textendash 13 in~\textbf{Algorithm~\ref{algorithm-1}} are asymptotically negligible. However, in a practical system, $N_{\mathrm{S}}\gg\max\{N_{\mathrm{T}},N_{\mathrm{R}},N_{\mathrm{U}},K,L\}$, therefore, the total computational complexity of the proposed AO-based scheme in~\textbf{Algorithm~\ref{algorithm-2}} is approximated by $\mathcal{O}\{KN_{\mathrm{U}}N_{\mathrm{T}}N_{\mathrm{S}}\}$, which is \emph{linear} in $N_{\mathrm{S}}$.

\section{Results and Discussion}

We consider a scenario where the BS is located at $(0,0,5)$~m, and the STAR-RIS is located at $(70,10,10)$~m. The communication users in the reflection and transmission regimes are randomly located inside a circle of radius $10$~m, centered at $(150,-20,2)$~m and $(150,40,2)$~m, respectively. The distance between $l$-th target and the BS is assumed to be equal to $(10+5l)$~m with an azimuth angle of $(20l)^{\circ}$. The direct links between the BS and communication users are modeled using Rayleigh fading with path loss exponent 3.6. The channel between the BS and STAR-RIS, and STAR-RIS and communication users follow Rician fading with Rician factor of $3$~dB, and path loss exponents equal to $2.2$ and $2.4$, respectively. The BS-target channels are modeled as line-of-sight (LoS) steering vectors with path loss exponent $2.2$. The path loss between two nodes at a distance of $d$~m is modeled as $[-30-10\varepsilon\log_{10}(d)]$~dB, where $\varepsilon$ is the path loss exponent. Furthermore, the self-interference link $\mathbf{G}$ is modeled following Rayleigh fading with variance equal to $\sigma^{2}$. 

For comparison of the system's performance with STAR-RIS, we consider two different benchmarks, namely conventional RIS (cRIS)-enabled system, and w/o RIS system. It is noteworthy that in the case of cRIS-enabled system, we assume that the system is equipped with a \emph{reflection-only} RIS of size $N_{\mathrm{S}}/2$, and a \emph{transmission-only} RIS of size $N_{\mathrm{S}}/2$, while for the case of w/o RIS system, we consider $\mathbf{H}_{\mathrm{BS}}=\boldsymbol{0}$ and $\mathbf{h}_{\mathrm{S}k}=\boldsymbol{0}\ \forall k\in\mathcal{K}$. Unless stated otherwise, in all the simulation results, we assume $N_{\mathrm{T}}=6$, $N_{\mathrm{R}}=4$, $N_{\mathrm{U}}=2$, $N_{\mathrm{S}}=100$, $K=4$ (with an equal number of users in the reflection and transmission regions of the STAR-RIS), $L=2$, $P_{\max}=10$~dBm, $\Delta_{\mathrm{s}l}=\ln(1+\Gamma_{\mathrm{s}})\ \forall l\in\mathcal{L}$, $\Gamma_{\mathrm{s}}=5$~dB, $\bar{\alpha}=0.5$, noise power spectral density of -174~dBm/Hz, and a system bandwidth of 10~MHz. Note that for Figs.~\ref{fig:vary_Ns}\textendash \ref{fig:vary_GammaS}, the average sum secrecy rate (ASSR) is obtained by averaging the instantaneous sum secrecy rate over 100 independent channel realizations. 

In Fig.~\ref{fig:convergence}, we show the convergence behavior of the STAR-RIS-enabled, cRIS-enabled and w/o RIS systems, where the augmented objective refers to $\mathcal{R}_{\boldsymbol{\nu},\rho}(\mathbf{J},\boldsymbol{\theta},\boldsymbol{\tau})$, and the true objective refers to $\sum_{k\in\mathcal{K}}R_{\mathrm{c}k}(\mathbf{J},\boldsymbol{\theta})$. We start the iterations with $\boldsymbol{\tau}=\boldsymbol{\nu}=\boldsymbol{0}$, $\rho=10$, and $\zeta=0.1$. It is important to note from~\eqref{eq:augmented_objective} that the augmented objective function $\mathcal{R}_{\boldsymbol{\nu},\rho}(\mathbf{J},\boldsymbol{\theta},\boldsymbol{\tau})$ may be negative during initial iterations, since $\mathscr{G}(\mathbf{J},\boldsymbol{\varphi}_{l},\Delta_{\mathrm{s}l},\tau_{l})\gg0$. This indicates that the sensing constraints in~\eqref{eq:sensing_rate-P1} are largely violated in initial iterations. It is also important to note that we compute the augmented and true objectives after step~13 of~\textbf{Algorithm~\ref{algorithm-1}}, and this corresponds to one iteration in the figure. Therefore, the objective function starts increasing and then saturates for fixed $(\boldsymbol{\nu},\rho)$ as shown in the figure. We then update $(\boldsymbol{\nu},\rho)$ following steps~3 and 4 in~\textbf{Algorithm~\ref{algorithm-2}}; this causes an increase in the total penalty, resulting in a sudden drop in the augmented objective value. We keep repeating this procedure until the convergence criterion for~\textbf{Algorithm~\ref{algorithm-2}} is achieved. At the convergence, the augmented objective becomes equal to the true objective, indicating $\mathscr{G}(\mathbf{J},\boldsymbol{\varphi}_{l},\Delta_{\mathrm{s}l},\tau_{l})=0$, i.e.,~\eqref{eq:sensing_rate-P1} being satisfied. Moreover, the rest of the constraints, i.e., \eqref{eq:TPC-P1}\textendash \eqref{eq:UMC-rxFiler-P1} are already satisfied in each iteration of~\textbf{Algorithm~\ref{algorithm-1}}. In~\textbf{Algorithm~\ref{algorithm-1}}, we define the convergence as the condition when $\{\mathcal R_{\boldsymbol{\nu}, \rho}(\mathbf J^{(\jmath)}, \boldsymbol{\theta}^{(\jmath)}, \boldsymbol{\tau}^{(\jmath)}) - \mathcal R_{\boldsymbol{\nu}, \rho}(\mathbf J^{(\jmath-5)}, \boldsymbol{\theta}^{(\jmath-5)}, \boldsymbol{\tau}^{(\jmath-5)})\} / \mathcal R_{\boldsymbol{\nu}, \rho}(\mathbf J^{(\jmath-5)}, \boldsymbol{\theta}^{(\jmath-5)}, \boldsymbol{\tau}^{(\jmath-5)}) \leq 10^{-5}$. In a similar fashion, we define the convergence for~\textbf{Algorithm~\ref{algorithm-2}} as $\{\mathcal R_{\boldsymbol{\nu}, \rho}(\mathbf J^{(\jmath)}, \boldsymbol{\theta}^{(\jmath)}, \boldsymbol{\tau}^{(\jmath)}) - \sum_{k\in \mathcal K} R_{\mathrm ck} (\mathbf J^{(\jmath-5)}, \boldsymbol{\theta}^{(\jmath-5)})\} / \mathcal R_{\boldsymbol{\nu}, \rho}(\mathbf J^{(\jmath)}, \boldsymbol{\theta}^{(\jmath)}, \boldsymbol{\tau}^{(\jmath)}) \leq 10^{-5}$.

Fig.~\ref{fig:vary_Ns} shows the impact of the number of metasurface elements ($N_{\mathrm{S}}$), and the number of transmit antennas at the BS ($N_{\mathrm{T}}$) on the system's performance. Note that since there are no metasurface elements in the case of w/o RIS system, the sum secrecy rate remains constant in the figure. It is evident from the figure that increasing $N_{\mathrm{S}}$ and/or $N_{\mathrm{T}}$ increases the ASSR of the MU-MIMO ISAC system due to increased beamforming gain. The figure also establishes the superiority of the STAR-RIS-aided system over that of the cRIS-aided and w/o RIS systems, in terms of the ASSR. More specifically, for $(N_{\mathrm{T}},N_{\mathrm{S}})=(6,16^{2})$, the performance gain of STAR-RIS-enabled system is $\approx92\%$ w.r.t. the w/o RIS, and $\approx59.3\%$ w.r.t the cRIS-enabled system. At the same time, for $N_{\mathrm{S}}=16^{2}$, we observe a $\approx73.3\%$ performance gain in STAR-RIS-enabled system by increasing $N_{\mathrm{T}}$ from 6 to 8.

In Fig.~\ref{fig:vary_Pmax}, we illustrate the impact of the BS transmit power budget ($P_{\max}$), the number of communication user antennas ($N_{\mathrm{U}}$), and the number of targets ($L$, also representing the number of eavesdropping antennas) on the STAR-RIS-enabled system. The legend's tuple denotes $(N_{\mathrm{U}}, L)$. For a given setup, a fixed portion of $P_{\max}$ is allocated to meet the sensing constraints in~\eqref{eq:sensing_rate-P1}, with the remaining power used for communication. Hence, increasing $P_{\max}$ enhances the available power for communication, boosting the ASSR. Similarly, higher $N_{\mathrm{U}}$ improves receive beamforming gain and further raises the sum secrecy rate. For instance, at $P_{\max}=20$~dBm and $L=1$, the sum secrecy rate improves by approximately $27.4\%$ for $N_{\mathrm{U}}=2$ and $57\%$ for $N_{\mathrm{U}}=4$, compared to $N_{\mathrm{U}}=1$. Conversely, as $L$ increases, the sensing constraints tighten, and the rise in eavesdropping antennas causes a significant decline in secrecy rate.

Fig.~\ref{fig:vary_GammaS} shows the impact of the sensing SINR threshold ($\Gamma_{\mathrm{s}}$) on the secrecy rate of the STAR-RIS-enabled system. As before, the legend's tuple denotes $(N_{\mathrm{U}}, L)$. Increasing $\Gamma_{\mathrm{s}}$ raises the transmit power needed to meet the sensing constraints in\eqref{eq:sensing_rate-P1}, leaving less power for communication and thus reducing the sum secrecy rate. When $L=1$ (i.e., a single-antenna eavesdropper), this reduction is marginal. However, as $L$ grows, the eavesdropper's spatial diversity becomes significant, leading to a sharp decline in the secrecy rate as $\Gamma_{\mathrm{s}}$ increases.

\section{Conclusion}
In this paper, we addressed the joint transmit, receive, and STAR-RIS beamforming design to maximize the sum secrecy rate in a MU-MIMO ISAC system, subject to sensing QoS for multiple targets, a transmit power budget, STAR-RIS power-splitting protocol, and unit-norm receive beamforming constraints. To achieve an efficient solution, we proposed an AO-based iterative algorithm, where receive beamformers are derived in closed-form, and transmit and STAR-RIS beamformers are computed using a penalty dual decomposition-based accelerated gradient projection method. Simulation results demonstrate that the STAR-RIS-enabled system outperforms both cRIS-enabled and no-RIS benchmarks. Moreover, while increasing the number of BS transmit antennas, metasurface elements, or communication receive antennas enhances the secrecy rate, a higher number of targets adversely affects system performance, particularly under stringent sensing QoS requirements.

\bibliographystyle{IEEEtran}
\bibliography{ref_MIMO-ISAC}

\end{document}